# Extension of the Injected-Absorbed-Current Method applied to DC-DC Converters with Input Filter, Output Post-filter and Feedforward Compensations

**Author 1:** Diego Ochoa / Universidad Carlos III de Madrid / Spain / dochoa@ing.uc3m.es

**Author 2:** Antonio Lázaro / Universidad Carlos III de Madrid / Spain / alazaro@ing.uc3m.es

**Author 3:** Pablo Zumel / Universidad Carlos III de Madrid / Spain / pzumel@ing.uc3m.es

**Author 4:** Cristina Fernández / Universidad Carlos III de Madrid / Spain / cfernand@ing.uc3m.es

**Author 5:** Marina Sanz / Universidad Carlos III de Madrid / Spain / cmsanz@ing.uc3m.es

**Author 6:** Jorge Rodríguez / Power Smart Control / Spain / jorge.rodriguez@powersmartcontrol.com

**Author 7:** Andrés Barrado / Universidad Carlos III de Madrid / Spain / andres.barrado@uc3m.es

ABSTRACT

In railway applications it is common to use an LC filter connected between the catenary and the input port of the main converter of the auxiliary and traction systems. In addition, in the auxiliary systems there is a converter operating as a battery charger, which requires a very low ripple in the output current and output voltage, so a post-filter may be placed at the output port of the converter. This paper proposes a step-by-step methodology to extend the injected-absorbed-current method in order to obtain transfer functions that take into account the effects of the input filter, output post-filter and some feedforward compensations. The proposed methodology allows reusing the characteristic coefficients of the DC-DC converter model derived from the existing injected-absorbed-current method. One of the advantages of the proposed methodology is that the transfer functions obtained in this paper are valid for cases where both, one or none of the filters are implemented. Finally, for the experimental validation of the proposed methodology, the Phase-Shifted Full-Bridge converter was selected as a convenient example. Furthermore, the experimental measurements have been performed on two prototypes.

*ABBREVIATIONS, ACRONYMS AND NOTATION*

| | |
|---|---|
| *IAC* | Injected-absorbed-current method. |
| *OIAC* | Original injected-absorbed-current method. |
| $EIAC$ | Extended injected-absorbed-current method. |
| $FF$ | Feedforward compensations. |
| $F_{ii}$ | Input current feedforward of the original converter without input filter. |
| $F_{vi}$ | Input voltage feedforward of the original converter without input filter. |
| $F_{ig}$ | Input current feedforward of the complete converter including the input filter. |
| $F_{vg}$ | Input voltage feedforward of the complete converter including the input filter. |
| $F_{io}$ | Output current feedforward of the complete converter including the output post-filter. |
| $G_m$ | Modulator transfer function. |
| $G_{sv}$ | Sensor transfer function. |
| $R_{eg}$ | Compensator transfer function. |
| $G_{vv}$ | Audio-susceptibility transfer function. |
| $G_{vvc}$ | Control to output voltage transfer function. |
| $Z_o$ | Output impedance transfer function. |
| $Z_{in}$ | Input impedance transfer function. |
| $A_x, B_x, C_x$ | Characteristic coefficients of the original injected-absorbed-current method. |
| $A'_x, B'_x, C'_x$ | Characteristic coefficients of the extended injected-absorbed-current method. |
| $G_x$ | Transfer functions. |
| $Z_x$ | Impedances. |
| $v_o^*$ | Output voltage reference. |
| $n$ | Transformer turns ratio (number of turns of the secondary winding divided by the number of turns of the primary winding). |
| $L_{lk}$ | Leakage inductance of the transformer. |
| $L$ | Output filter inductance. |

# I. INTRODUCTION

Nowadays, the power demand on the converters of auxiliary and traction systems in railway applications is in the range of hundred-kilowatts or Megawatts. The use of an isolated converter is required for safety reasons. In recent years, the use of DC electrification rather than AC electrification is preferred in railway applications [1]-[2], as having an isolated DC-DC converter allows the switching frequency to be increased up to several kilohertz, which leads to a reduction in the size and weight of the transformers [3]-[4]. In addition, by using topologies that allow the implementation of some soft-switching technique, switching power losses can be reduced [4]-[6].

A typically structure of auxiliary system in railway applications is shown in Fig. 1. The most common catenary voltages in Europe are 750 Vdc, 1500 Vdc and 3000 Vdc [7]. The main isolated DC-DC converter reduces the catenary voltage to a bus voltage, $v_{bus}$, of 500-700 Vdc. This converter is exposed to very high voltage levels, so it is very common for the main DC-DC converter to be implemented in an input-series-output-parallel (ISOP) architecture or any other modular architecture in order to use switching devices that can support lower voltages [6], [8]-[9]. On the other hand, the converter that operates as a battery charger reduces the bus voltage to a typical voltage of 36/72/110 Vdc [10]. This converter is essential in railway applications as it is responsible for supplying power to all control boards, gate drivers and TCUs [10]. Finally, the three-phase inverter converts the bus voltage to 380 Vac [8]. This converter is responsible for supplying power to the AC loads such as the lighting system, air conditioning compressors and passenger information systems.

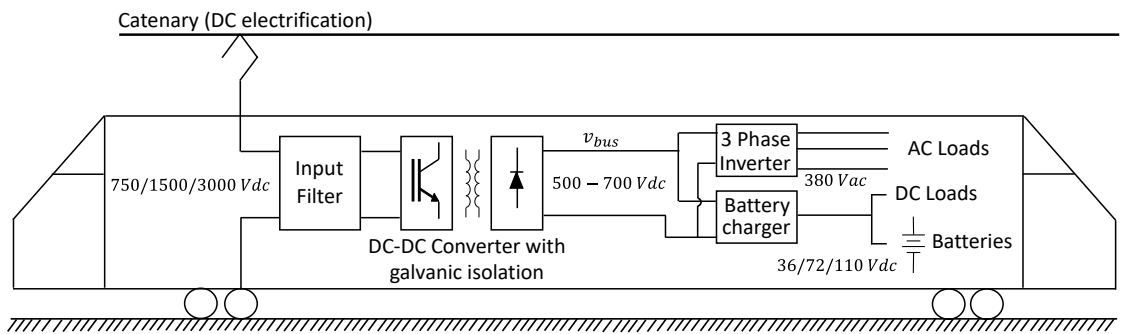

Fig. 1 Auxiliary system in railway applications

The catenary voltage is exposed to fast changes due to the connection and disconnection of the pantograph in cross-section changes and gaps. So, the input LC filter is required in order to filter voltage spikes, attenuate electromagnetic interference (EMI) and comply with the input impedance requirement specified by some regulations or standards [11]-[12].

On the other hand, a second LC filter or post-filter may be placed at the output of the battery charger converter in order to reduce the output current ripple or output voltage ripple. This post-filter is especially significative in those converters that deliver pulsating current to the output filter capacitor (e.g., boost, buck-boost, resonant converters, and dual active bridge), since a single stage filter could lead to an oversizing of the filter capacitor and could be more expensive [13]-[15].

The dynamic interaction between the input filter and the regulated converter connected at its output is very well known [16] and it is related to the constant power load (CPL) nature of the regulated converter. The incremental negative impedance that is present in constant power loads (CPL) also degrades the stability of cascade converters [17]. A review of the main stability criteria is shown in [17] and the fundamental idea is focused on the analysis of the minor loop gain (TMLG) that is defined as the relationship between the output impedance of the input filter or source converter and the input impedance of the load converter [17]. The TMLG is the responsible for the stability when two stable subsystems are interconnected.

The main elements found in railway applications such as input filter, feedforward compensations and constant power loads, are shown in Fig. 2. These three elements justify the need to obtain transfer functions that allow analyzing the stability criterion based on the TMLG and to design optimal control loops.

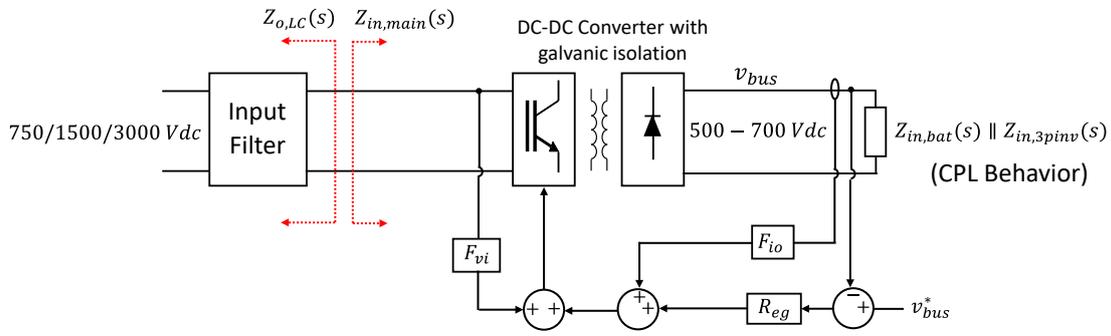

Fig. 2 Simplified scheme of the auxiliary systems that represent the basic elements that are present in railway applications

A brief analysis of how the methodology proposed in this paper takes into account the constant power load behavior is provided. First the closed-loop input impedance of the battery charger, $Z_{in,bat}$, and closed-loop input impedance of the three-phase inverter, $Z_{in,3pinv}$, should be calculated. The input impedance of the battery charger must take into account the effect of the output post-filter and feedforward compensations. A small-signal stability analysis is considered in this paper, so it is assumed that the source and load converter can be represented by small-signal models. The resulting input impedance of the downstream converters is defined by (1).

$$Z_{in,load}(s) = Z_{in,bat}(s) \parallel Z_{in,3pinv}(s) \qquad (1)$$

The control to output voltage transfer function of the main DC-DC converter must consider the expression (1) as the load as shown in Fig. 2, since this takes into account the constant power load behavior (CPL) of the downstream converters. In addition, the input filter must also be considered in the control to output voltage transfer function in order to design an optimal control loop.

Once the compensator of the main DC-DC converter has been tuned, the TMLG must be analyzed to verify that the stability criterion is still met. The TMLG is defined by the relationship between the output impedance of the LC input filter, $Z_{o,LC}(s)$, and the input impedance of the main converter, $Z_{in,main}(s)$.

$$TMLG = \frac{Z_{o,LC}(s)}{Z_{in,main}(s)} \qquad (2)$$

From this analysis, it can be seen that it is necessary to calculate transfer functions that take into account the effects of the input filter, output post-filter and feedforward compensations with the aim of ensuring the stability of the whole system shown in Fig. 1.

The Feedforward compensations are considered in the proposed methodology for the following reasons. The closed-loop input and output impedance depend on the control technique, so there are strategies known as active damping stabilization whose function is to modify the input or output impedance with the purpose that the minor loop gain (TLMG) meets the stability criteria. Many of the active damping techniques utilizes the positive feedforward compensation in addition to the conventional negative feedback control in order to modify the closed-loop input impedance (input voltage feedforward compensation and input current feedforward compensation) [18]-[22] or in order to modify the closed-loop output impedance (output voltage feedforward compensation and output current feedforward compensation) [23]. On the other hand, the use of an input voltage feedforward compensation (IVFF) is also a common practice in applications with wide input voltage variation such as railway applications [24]. The IVFF rejects the perturbation of the input voltage over the output voltage or output current in consequence improves the transient response of the control loop [25]-[27]. In the same way, the output current feedforward (OCFF) improves the dynamic response under fast load changes, since OCFF rejects the perturbation of the load current [28]-[30].

Determining transfer functions that consider the input filter, output post-filter and FF compensations is a complex and very time-consuming task. So, a simple methodology is proposed together with a new set of general or canonical transfer functions that can be applied to almost any power converter and control technique.

The average modeling techniques allow to obtain a good approximation of the dynamic response of the power converter up to half of the switching frequency. Some of the most popular average modeling techniques to obtain generic transfer functions are: the state-space averaging method (SSA) and the injected-absorbed-current method (IAC). In the SSA method [31]-[32], the small-signal model is deduced from the matrices that represent the differential equations of the state variables. When both filters are implemented, the matrix operation of SSA method can become quite complicated, since the number of state variable are notoriously increased. On the other hand, the IAC method [33]-[35] allows to represent the small-signal model of the any converter topology by means of a generalized electrical equivalent circuit. This equivalent circuit depends on the six characteristic coefficients that relates the input or absorbed current and the output or injected current of the converter with the input voltage, output voltage and control quantity.

Since both methods are linear, the results obtained from the IAC method and the SSA method are equivalent. However, the circuit model resulted from the SSA method varies according to the conduction mode (continuous or discontinuous conduction mode) and if the controlled quantity is other than duty ratio [36]-[37]. On the other hand, the circuit model resulted from the IAC method is always the same and is valid for all type of converters [38] that can have constant-frequency or free-running configuration in continuous or discontinuous conduction mode with the controlled quantity being duty ratio or switching frequency [34], [39]. Therefore, the IAC method has been chosen as a model technique, since it is the most convenient technique to determine certain transfer functions such as input and output impedance in an effective way. The SSA method can be used as a previous step to obtain the characteristic coefficients of the IAC method [40].

The IAC method was presented in [33] and is extended in [34]-[35] to some particular cases where the converter has an input filter, $F_{vi}$, $F_{ig}$ and $F_{io}$. However, these references do not provide the complete and general expressions of the converter transfer functions when the converter has an input filter, $F_{vi}$, $F_{ig}$ and $F_{io}$ at the same time.

On the other hand, at this time no references can be found in the literature regarding to the extension of IAC method to situations in which an output post-filter is added to the power converter.

Therefore, this paper complements the references [34]-[35] to general situations where different types of filters (input filter and output post-filter) and the different types of feedforward compensations can be implemented. The FF to consider in the EIAC method are $F_{vi}$, $F_{ig}$, $F_{io}$ and even two new feedforward compensations are considered that are $F_{ii}$ and $F_{vg}$.

Since obtaining transfer functions of the complete circuit is a very hard task, then in this paper a step-by-step methodology is proposed to extend the IAC method in order to obtain transfer functions that take into account the effects of the input filter, output post-filter and some feedforward compensations. The main advantage of this extension is that due to the similarity with the classical IAC circuit, the generic transfer functions obtained in this paper such as closed-loop input impedance, closed-loop output impedance, closed-loop audio-susceptibility, closed-loop back current and control to output voltage are valid for cases where both, one or none of the filters are implemented. The proposed methodology allows reusing the characteristic coefficients of the DC-DC converter model derived from the existing injected-absorbed current method. The Characteristic coefficients of some converters can be found in the existing literature, e.g., Buck converter, Boost converter, Buck-Boost converter, Cuk converter, SEPIC converter, Series resonant converter, Parallel resonant converter, and Phase-Shifted Full-bridge converter [34], [41]-[45]. The most used topologies in the railway industry are the Phase-Shifted Full-Bridge converter and series resonant converter [24], [46]-[52].

The Phase-Shifted Full-Bridge (PSFB) converter [53] is chosen in order to validate the transfer functions obtained from the extension of the IAC method, as it is one of the most popular converters for medium-high power applications, due to its zero-voltage-switching (ZVS) capability, high efficiency, fixed switching frequency due to its PWM modulation, and its naturally reduced output voltage ripple (it is a buck derived topology). In the PSFB converter the transformer leakage inductance, $L_{lk}$, plays an important role in achieving ZVS operation, since the energy stored in the leakage inductance must be able to discharge the output capacitor of the MOSFET/IGBT in order to force the conduction of the antiparallel diode prior to the conduction of the MOSFET/IGBT. The voltage drop in this leakage inductance creates a blanking time interval, which leads to a reduction across the pulse width of the voltage reflected at the secondary of the transformer. This interval is not a controllable quantity and depends only on the operating conditions of the converter.

The original contributions of this paper are:

- The methodology to extend the IAC method that allows obtaining general converter transfer functions and impedances, that can be applied to any converter topology with input filter, output post-filter and feedforward compensations of the input voltage, input current and output current.
- A new set of converter transfer functions and impedances, such as control to output voltage, closed-loop input impedance, closed-loop output impedance (terminated and unterminated cases), closed-loop back current and closed-loop audio-susceptibility are obtained as a final step of the extension of IAC method.

The paper is organized as follows. In Section II, the methodology to extend the injected-absorbed current method is shown. In Section III, validation of the extension of the IAC method is performed. Finally, conclusions are drawn in Section IV.

## II. METHODOLOGY TO EXTEND THE INJECTED-ABSORBED-CURRENT METHOD

### A. Introduction of the extension of the IAC method

The injected-absorbed-current method allows to represent any power converter topology in which the average value of the output current (injected current) and the average value of the input current (absorbed current) can be expressed as function of the average value of the input voltage, output voltage and control quantity [34]. The control quantity is the parameter that allows controlling the energy transfer in the power converter and can be the duty ratio (at constant or variable frequency) or the frequency (at constant duty ratio). The validity of the injected-absorbed-current method is limited to low-frequency phenomena and small signals.

In this paper, the expressions "original IAC model" refers to the power converter without considering any external elements, "power structure" refers to the different combinations that can be given by the filters, "control structure" refers to the different combinations that can be given by the feedforward compensations and "extended IAC model" refers to the IAC model of the power converter including the input filter, output post filter and feedforward compensations.

The most critical scenario is when the converter has an input filter, output post-filter and feedforward compensations at the same time as shown in Fig 3. This scenario could be the case of the converter that operates as a battery charger. The feedforward compensations of the control structure are classified as internal and external. The input voltage feedforward compensation, $F_{vi}(s)$, and input current feedforward compensation, $F_{ii}(s)$, of the original converter without input filter are part of the internal feedforward compensations and the input current feedforward compensation, $F_{ig}(s)$, and input voltage feedforward compensation, $F_{vg}(s)$, of the complete converter including the input filter and the output current feedforward compensation, $F_{io}(s)$, of the complete converter including the output post-filter are part of the external feedforward compensations.

Due to the complexity of deducing transfer functions from the circuit shown in Fig. 3, in this paper a methodology to extend the IAC method is proposed in order to deduce generic transfer functions that can be applied to different topologies with different power and control structures and thus simplify the modeling process.

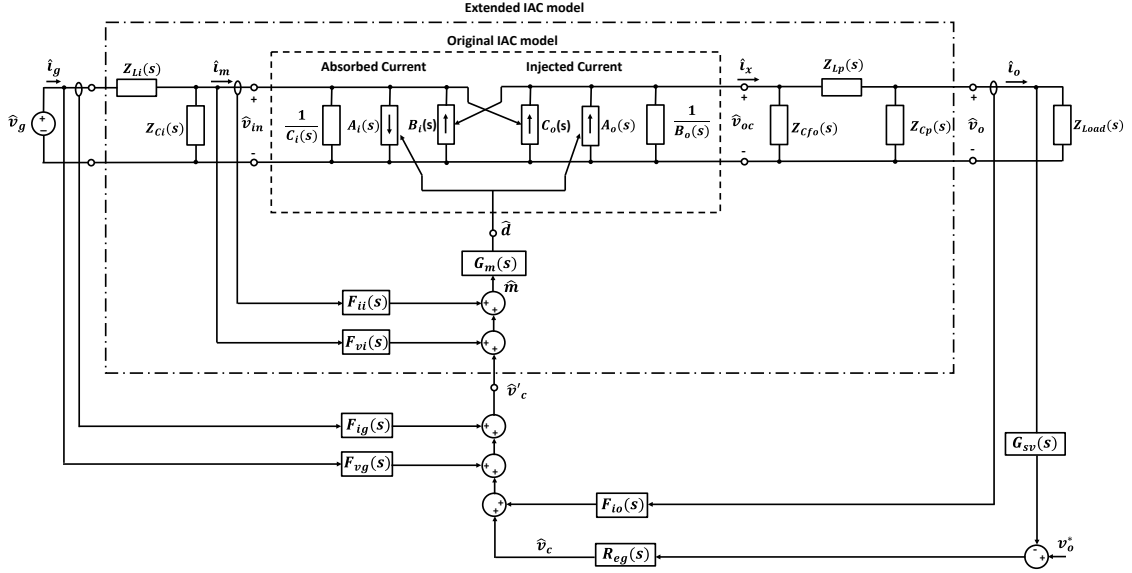

Fig. 3 Complete power and control structures of the converter

B. Methodology to extend the IAC method

The proposed methodology consists of three steps: The first step is to obtain the characteristic coefficients of the original IAC model. The characteristic coefficients represent the dynamics of the input current, $\hat{\imath}_m$, and the output current, $\hat{\imath}_x$, of the power converter when there are perturbations in the input voltage, $\hat{v}_{in}$, output voltage, $\hat{v}_{oc}$, and control quantity, $\hat{d}$. The controlled quantity for this analysis is the duty ratio.

The small-signal circuit resulting from the IAC method is shown in Fig 4. The absorbed current and the injected current are given by (3) and (4) respectively. The characteristic coefficients of the output port are $A_o(s)$, $B_o(s)$ and $C_o(s)$. The characteristic coefficients of the input port are $A_i(s)$, $B_i(s)$ and $C_i(s)$.

$$\hat{\imath}_m = A_i(s) \cdot \hat{d} - B_i(s) \cdot \hat{v}_{oc} + C_i(s) \cdot \hat{v}_{in} \qquad (3)$$

$$\hat{\imath}_x = A_o(s) \cdot \hat{d} - B_o(s) \cdot \hat{v}_{oc} + C_o(s) \cdot \hat{v}_{in} \qquad (4)$$

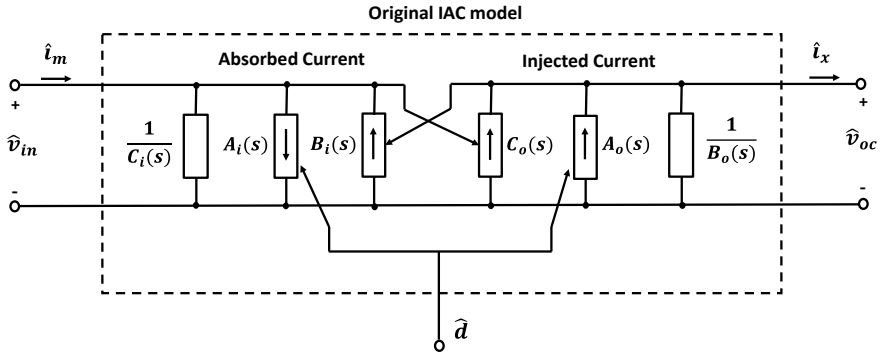

Fig. 4 Original injected-absorbed-current model

The second step of the proposed methodology is to take the circuit shown in Fig. 5(a) to a simplified circuit as shown in Fig. 5(b).

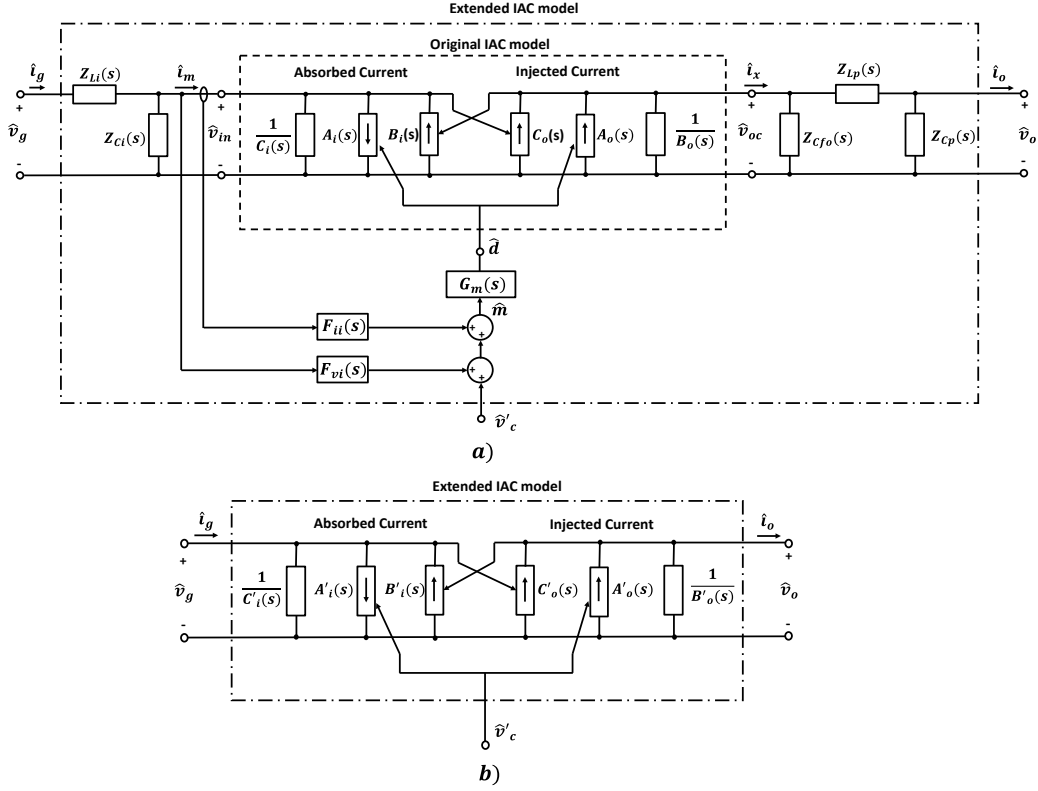

Fig. 5 a) IAC model with input filter, output post-filter and internal feedforward compensations b) EIAC model

It can be seen from Fig. 5(b) that the absorbed current, $\hat{\imath}_g$, and the injected current, $\hat{\imath}_o$, of the EIAC model depend on the small-signal magnitudes $\hat{v}'_c$, $\hat{v}_o$ and $\hat{v}_g$ and they are given by (5) and (6) respectively. These new characteristic coefficients include the effect of input filter, output post-filter and internal feedforward compensations.

$$\hat{\imath}_g = A'_i(s) \cdot \hat{v}'_c - B'_i(s) \cdot \hat{v}_o + C'_i(s) \cdot \hat{v}_g \qquad (5)$$

$$\hat{\imath}_o = A'_o(s) \cdot \hat{v}'_c - B'_o(s) \cdot \hat{v}_o + C'_o(s) \cdot \hat{v}_g \qquad (6)$$

In this step it is necessary to find the relationship that exists between the new characteristic coefficients that describes the EIAC model and the characteristic coefficients that describe the original IAC model.

In order to broaden the scope of the proposed methodology, four possible power structures have been developed:

- Power structure 1: Original power converter with input filter and output post-filter.
- Power structure 2: Original power converter with input filter.
- Power structure 3: Original power converter with output post-filter.

• Power structure 4: Original power converter without filters.

This classification has been developed since the new characteristic coefficients of power structures 2, 3 and 4 cannot be inferred without complex algebraic manipulations from the new characteristic coefficients of power structure 1. In this paper the characteristic coefficients of the EIAC model are determined for the four power structures mentioned above.

The main advantage of this extension is that the process for deriving the converter transfer functions becomes easier. In addition, the transfer functions that have been obtained in previous works from the OIAC model can be reusable in the case of the EIAC model thanks to the similarity that exists between the circuit shown in Fig. 4 and the circuit shown in Fig. 5(b).

The third step is to deduce the transfer functions from the circuit shown in Fig. 6. This paper focuses on single-loop voltage mode control but can be easily extended to current mode control, peak current mode control and double loop control or average current mode control.

Fig. 6 EIAC model under voltage mode control

The transfer functions obtained in this paper are the control to output voltage, closed-loop input impedance, closed-loop output impedance without considering the load, unterminated closed-loop back-current and closed-loop audio-susceptibility. All these transfer functions take into account internal and external feedforward compensations.

## C. Complete derivation of the new characteristic coefficients that describe the EIAC model

The main objective of this section is to calculate the new characteristic coefficients of the equivalent model shown in Fig. 5(b) for the different power structures.

The power structure 1 is shown in Fig. 5(a). Note in expressions (5) and (6) that the absorbed current, $\hat{i}_g$, and the injected current, $\hat{i}_o$, of the EIAC model depend on the quantities of the $\hat{v}_g$, $\hat{v}_o$ and the new control quantity $\hat{v}'_c$. Therefore, it is necessary to find the relationship that exists between the new characteristic coefficients that describes the EIAC model and the characteristic coefficients that describe the original IAC model.

From Fig. 5(a), one can obtain (7) and (8).

$$\hat{i}_m = \frac{\hat{v}_g}{Z_{Li}(s)} - \frac{\hat{v}_{in}}{Z_g(s)} \quad (7)$$

$$Z_g(s) = Z_{Li}(s) \parallel Z_{Ci}(s) \quad (8)$$

Similarly, a relationship between the voltages $\hat{v}_{oc}$ and $\hat{v}_o$ is obtained.

$$\hat{i}_o = \frac{\hat{v}_{oc}}{Z_{Lp}(s)} - \frac{\hat{v}_o}{Z_{op}(s)} \quad (9)$$

$$Z_{op}(s) = Z_{Lp}(s) \parallel Z_{Cp}(s) \quad (10)$$

$$\hat{i}_x = \frac{\hat{v}_{oc}}{Z_{LpC}(s)} - \frac{\hat{v}_o}{Z_{Lp}(s)} \quad (11)$$

$$Z_{LpC}(s) = Z_{Lp}(s) \parallel Z_{Cfo}(s) \quad (12)$$

The control quantity for the power structure 1 is given by (13).

$$\hat{d} = G_m(s) \cdot [\hat{v}'_c + F_{ii}(s) \cdot \hat{i}_m + F_{vi}(s) \cdot \hat{v}_{in}] \quad (13)$$

Using (3), (4), (7), (9), (11) and (13), the absorbed current, $\hat{i}_g$, and the injected current, $\hat{i}_o$, can be obtained as functions of the small increments of the $\hat{v}_g$, $\hat{v}_o$ and $\hat{v}'_c$. Table I shows the new characteristic coefficients that describe the expressions (5) and (6). For this power architecture the term $B_o(s)$ should not include the output filter capacitor impedance, $Z_{cfo}(s)$. The new characteristic coefficients are functions of the complex variable "s". In Table I to V, the dependence on the Laplace variable has been omitted for simplicity. Note that the internal feedforward compensations are within the new characteristic coefficients. The external feedforward compensations will be considered as part of the converter transfer functions.

TABLE I

CHARACTERISTIC COEFFICIENTS CONSIDERING INPUT FILTER, OUTPUT POST-FILTER AND INTERNAL FEEDFORWARD COMPENSATIONS

| | |
|---|---|
| $x_1 = C_i - \dfrac{A_i \cdot F_{ii} \cdot G_m - 1}{Z_g} + A_i \cdot F_{vi} \cdot G_m$ | $x_2 = A_o \cdot F_{vi} \cdot G_m - \dfrac{A_o \cdot F_{ii} \cdot G_m}{Z_g} + C_o$ |

$$A'_o = \frac{A_o \cdot G_m - \frac{G_m \cdot A_i \cdot x_2}{x_1}}{\frac{Z_{Lp}}{Z_{LpC}} - \frac{Z_{Lp} \cdot B_i \cdot x_2}{x_1} + B_o \cdot Z_{LP}}$$

$$B'_o = \frac{\frac{Z_{Lp}}{Z_{op}} \cdot \left(B_o + \frac{1}{Z_{LpC}}\right) - \frac{1}{Z_{Lp}} - \frac{B_i \cdot Z_{Lp} \cdot x_2}{Z_{op} \cdot x_1}}{\frac{Z_{Lp}}{Z_{LpC}} - \frac{B_i \cdot Z_{Lp} \cdot x_2}{x_1} + B_o \cdot Z_{LP}}$$

$$C'_o = \frac{\frac{G_m \cdot A_o \cdot F_{ii}}{Z_{Li}} + \frac{(1 - A_i \cdot F_{ii} \cdot G_m) \cdot x_2}{x_1 \cdot Z_{Li}}}{\frac{Z_{Lp}}{Z_{LpC}} - \frac{Z_{Lp} \cdot B_i \cdot x_2}{x_1} + Z_{Lp} \cdot B_o}$$

$$B'_i = \frac{\frac{B_i \cdot Z_{Lp}}{Z_{op} \cdot Z_{Ci}} \cdot (Z_g - Z_{Ci}) \cdot (B'_o \cdot Z_{op} - 1)}{Z_g \cdot C_i + A_i \cdot G_m \cdot (F_{vi} \cdot Z_g - F_{ii}) + 1}$$

$$A'_i = \frac{(Z_{Ci} - Z_g) \cdot (A_i \cdot G_m - A'_o \cdot B_i \cdot Z_{Lp})}{Z_{Ci} \cdot (Z_g \cdot C_i - A_i \cdot F_{ii} \cdot G_m + A_i \cdot F_{vi} \cdot G_m \cdot Z_g + 1)}$$

$$C'_i = \frac{Z_g + C_i \cdot Z_g \cdot Z_{Ci} + A_i \cdot G_m \cdot Z_g \cdot (F_{vi} \cdot Z_{Ci} - F_{ii}) + B_i \cdot C'_o \cdot Z_{Li} \cdot Z_{Lp} \cdot (Z_g - Z_{Ci})}{Z_{Li} \cdot Z_{Ci} \cdot (Z_g \cdot C_i - A_i \cdot F_{ii} \cdot G_m + A_i \cdot F_{vi} \cdot G_m \cdot Z_g + 1)}$$

The power structure 2 is shown in Fig. 7. Note that if the term $B_o(s)$ considers the output filter capacitor impedance, $Z_{Cfo}(s)$, then one can obtain (14) and (15). The control quantity is given by (13).

$$\hat{i}_o = \hat{i}_x \qquad (14)$$

$$\hat{v}_o = \hat{v}_{oc} \qquad (15)$$

Using (3), (4), (7), (13), (14) and (15), the absorbed current, $\hat{i}_g$, and the injected current, $\hat{i}_o$, can be obtained as functions of the small increments of the $\hat{v}_g$, $\hat{v}_o$ and $\hat{v}'_c$. Table II shows the resulting characteristic coefficients of the EIAC model that is in Fig. 5(b). Since the resulting EIAC model of power structure 2 is the same as power structure 1, expressions (5) and (6) are still valid. This power structure is the one found in the main converter of auxiliary systems in railway applications.

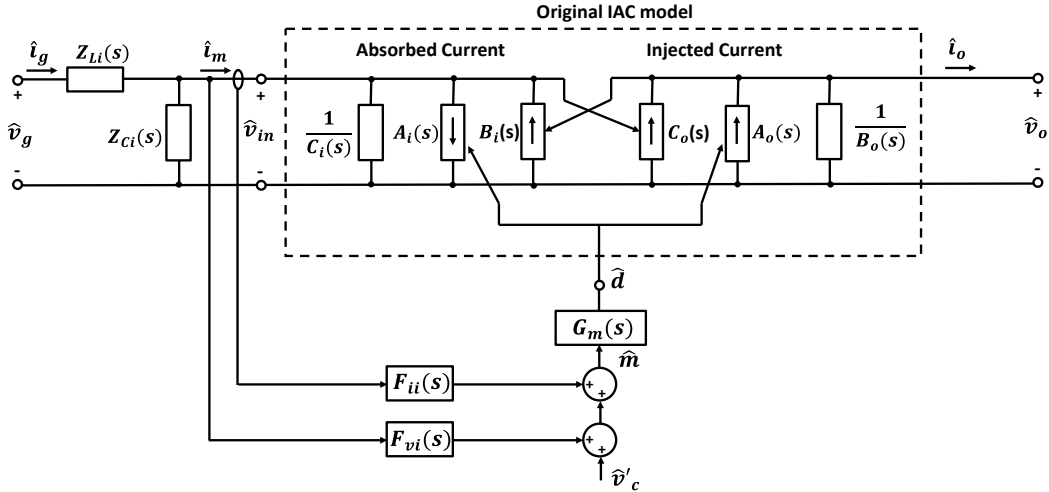

Fig. 7 Complete scheme of power structure 2

TABLE II

CHARACTERISTIC COEFFICIENTS CONSIDERING INPUT FILTER AND INTERNAL FEEDFORWARD COMPENSATIONS

| | |
|---|---|
| $A'_o = \dfrac{G_m \cdot (A_o - A_i \cdot C_o \cdot Z_g + A_o \cdot C_i \cdot Z_g)}{C_i \cdot Z_g - A_i \cdot G_m \cdot (F_{ii} - F_{vi} \cdot Z_g) + 1}$ | $B'_o = B_o - \dfrac{B_i \cdot C_o \cdot Z_g - A_o \cdot B_i \cdot F_{ii} \cdot G_m + A_o \cdot B_i \cdot F_{vi} \cdot G_m \cdot Z_g}{C_i \cdot Z_g - A_i \cdot F_{ii} \cdot G_m + A_i \cdot F_{vi} \cdot G_m \cdot Z_g + 1}$ |
| $C'_o = \dfrac{C_o + G_m \cdot (A_o \cdot (F_{vi} + C_i \cdot F_{ii}) - A_i \cdot C_o \cdot F_{ii})}{\dfrac{Z_{Li}}{Z_g} \cdot (C_i \cdot Z_g - A_i \cdot G_m \cdot (F_{ii} - F_{vi} \cdot Z_g) + 1)}$ | $A'_i = \dfrac{A_i \cdot G_m \cdot (Z_{Ci} - Z_g)}{Z_{Ci} \cdot (C_i \cdot Z_g - A_i \cdot G_m \cdot (F_{ii} - F_{vi} \cdot Z_g) + 1)}$ |
| $B'_i = \dfrac{B_i \cdot (Z_{Ci} - Z_g)}{Z_{Ci} \cdot (C_i \cdot Z_g - A_i \cdot G_m \cdot (F_{ii} - F_{vi} \cdot Z_g) + 1)}$ | $C'_i = \dfrac{Z_g \cdot (C_i \cdot Z_{Ci} - A_i \cdot F_{ii} \cdot G_m + A_i \cdot F_{vi} \cdot G_m \cdot Z_{Ci} + 1)}{Z_{Li} \cdot Z_{Ci} \cdot (C_i \cdot Z_g - A_i \cdot G_m \cdot (F_{ii} - F_{vi} \cdot Z_g) + 1)}$ |

The power structure 3 is shown in Fig. 8. For this case, the expressions (9), (10), (11) and (12) are valid. Since there is no input filter in this structure, then $F_{ii}$ is equal to $F_{ig}$ and $F_{vi}$ is equal to $F_{vg}$. Therefore, the control quantity for power structure 3 is given by (16).

$$\hat{d} = G_m(s) \cdot \hat{v}'_c \qquad (16)$$

$$\hat{\imath}_g = \hat{\imath}_m \qquad (17)$$

$$\hat{v}_g = \hat{v}_{in} \qquad (18)$$

Using (3), (4), (9), (11), (16), (17) and (18) the absorbed current, $\hat{\imath}_g$, and the injected current, $\hat{\imath}_o$, can be obtained as functions of the small increments of the $\hat{v}_g$, $\hat{v}_o$ and $\hat{v}'_c$. Table III shows the resulting characteristic coefficients of the EIAC model that is in Fig. 5(b). The expressions (5) and (6) are still valid. For this power architecture the term $B_o(s)$ should not include the output filter capacitor impedance, $Z_{cfo}(s)$. In order to maintain the same nomenclature in the EIAC model, it is assumed that $\hat{m}$ is equal to $\hat{v}'_c$ in this power structure.

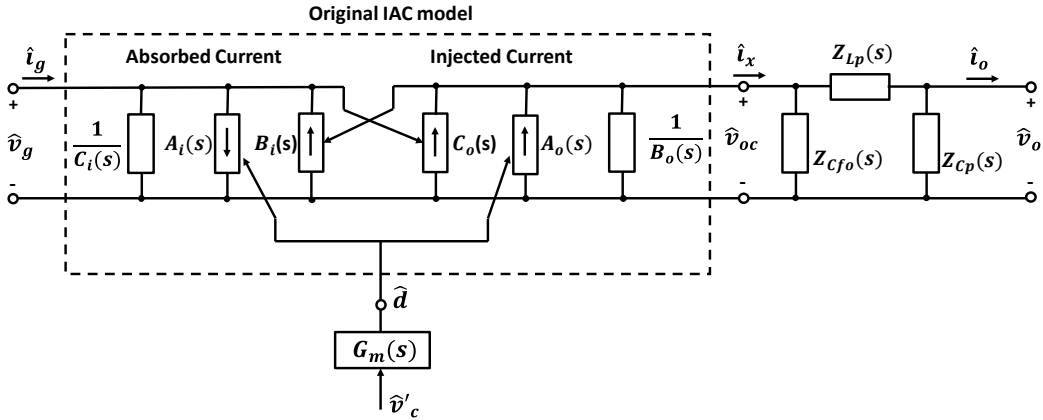

Fig. 8 Complete scheme of power structure 3

TABLE III

CHARACTERISTIC COEFFICIENTS CONSIDERING OUTPUT POST- FILTER

| $A'_o = \dfrac{A_o \cdot G_m \cdot Z_{LpC}}{Z_{LP} \cdot (B_o \cdot Z_{LpC} + 1)}$ | $B'_o = \dfrac{Z_{Lp}^2 - Z_{op} \cdot Z_{LpC} + B_o \cdot Z_{Lp}^2 \cdot Z_{LpC}}{Z_{Lp}^2 \cdot Z_{op} \cdot (B_o \cdot Z_{LpC} + 1)}$ | $C'_o = \dfrac{C_o \cdot Z_{LpC}}{Z_{LP} \cdot (B_o \cdot Z_{LpC} + 1)}$ |
|---|---|---|
| $A'_i = A_i \cdot G_m - \dfrac{A_o \cdot B_i \cdot G_m \cdot Z_{LpC}}{B_o \cdot Z_{LpC} + 1}$ | $B'_i = \dfrac{B_i \cdot Z_{LpC}}{Z_{LP} \cdot (B_o \cdot Z_{LpC} + 1)}$ | $C'_i = C_i - \dfrac{C_o \cdot B_i \cdot Z_{LpC}}{B_o \cdot Z_{LpC} + 1}$ |

Finally, the power structure 4 is shown in Fig. 9. Note that if the term $B_o(s)$ considers the output filter capacitor impedance, $Z_{Cfo}(s)$, then expressions (14) and (15) are valid. On the other hand, in this power structure, since it does not have an input filter, the expressions (17) and (18) are valid. The control quantity is given by (16).

Using (3), (4), (14), (15), (16), (17) and (18), the absorbed current, $\hat{\imath}_g$, and the injected current, $\hat{\imath}_o$, can be obtained as functions of the small increments of the $\hat{v}_g$, $\hat{v}_o$ and $\hat{v}'_c$. Table IV shows the resulting characteristic coefficients of the EIAC model that is in Fig. 5(b). The expressions (5) and (6) are still valid

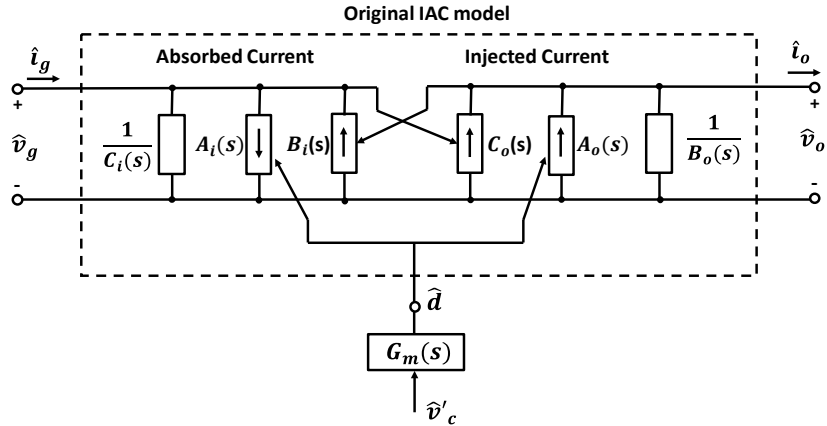

Fig. 9 Complete scheme of power structure 4

TABLE IV

CHARACTERISTIC COEFFICIENTS WITHOUT CONSIDERING BOTH FILTERS

| $A'_o = A_o \cdot G_m$ | $B'_o = B_o$ | $C'_o = C_o$ |
|---|---|---|
| $A'_i = A_i \cdot G_m$ | $B'_i = B_i$ | $C'_i = C_i$ |

The EIAC model is the same for the four power structures, so the main advantage of this strategy is that the transfer functions obtained in the third step of the proposed methodology are valid for the four power structures, since these transfer functions depend on the new characteristic coefficients that are specified in Table I, Table II, Table III and Table IV.

D. Complete derivation of the converter transfer functions

This paper focuses on single loop voltage-mode control but can be easily extended to current mode control, peak current mode control and double loop control or average current mode control. The control structure from which transfer functions are obtained is shown in Fig. 6. In the control structure, the modulator transfer function, $G_m$, is not considered, since it is already considered within the new characteristic coefficients.

To obtain the control to output voltage transfer function, $G_{vvc}(s)$, the control quantity is given by (19).

$$\hat{v}'_c = \hat{v}_c + F_{io}(s) \cdot \hat{\imath}_o + F_{vg}(s) \cdot \hat{v}_g + F_{ig}(s) \cdot \hat{\imath}_g \tag{19}$$

From Fig. 6 it follows that the output current is given by (20).

$$\hat{\imath}_o = \frac{\hat{v}_o}{Z_{load}(s)} \tag{20}$$

The term $Z_{load}(s)$ represents the load impedance. The EIAC model is valid for different types of loads such as resistive load, constant power load, etc. In the introduction section it is mentioned how to use the model in case of constant power loads.

On the other hand, to obtain the closed-loop transfer functions, the control signal is defined by (21).

$$\hat{v}'_c = -G_{sv}(s) \cdot R_{eg}(s) \cdot \hat{v}_o + F_{io}(s) \cdot \hat{\imath}_o + F_{vg}(s) \cdot \hat{v}_g + F_{ig}(s) \cdot \hat{\imath}_g \tag{21}$$

Using (5), (6), (19), (20) and (21), the different transfer functions shown in Table V can be obtained. Appendix A provides step-by-step details on how to obtain the transfer functions found in Table V.

TABLE V

CONVERTER TRANSFER FUNCTIONS

| Control to output voltage transfer function, $G_{vvc}(s)$ |
|---|
| $\dfrac{\hat{v}_o}{\hat{v}_c} = \dfrac{A'_o \cdot Z_{load}}{F_{ig} \cdot Z_{load} \cdot (A'_o \cdot B'_i - A'_i \cdot B'_o) + 1 + B'_o \cdot Z_{load} - A'_o \cdot F_{io} - A'_i \cdot F_{ig}}$ |
| Closed-loop input impedance transfer function, $Z_{in}(s)$ |
| $\dfrac{\hat{v}_g}{\hat{\imath}_g} = \dfrac{B'_o - \dfrac{A'_o \cdot F_{io}}{Z_{load}} - \dfrac{A'_i \cdot F_{ig}}{Z_{load}} - A'_i \cdot B'_o \cdot F_{ig} + A'_o \cdot B'_i \cdot F_{ig} + A'_o \cdot G_{sv} \cdot R_{eg} + \dfrac{1}{Z_{load}}}{\dfrac{C'_i}{Z_{load}} + \dfrac{A'_i \cdot F_{vg}}{Z_{load}} - C'_o \cdot B'_i + C'_i \cdot B'_o + \dfrac{F_{io}}{Z_{load}} \cdot (A'_i \cdot C'_o - A'_o \cdot C'_i) + F_{vg} \cdot (A'_i \cdot B'_o - A'_o \cdot B'_i) - G_{sv} \cdot R_{eg} \cdot (A'_i \cdot C'_o - A'_o \cdot C'_i)}$ |
| Unterminated closed-loop output impedance transfer function (without considering the load), $Z_{o,un}(s)$ |
| $Z_{o,un}(s) = \dfrac{\hat{v}_o}{\hat{\imath}_o} = \dfrac{A'_i \cdot F_{ig} + A'_o \cdot F_{io} - 1}{B'_o - A'_i \cdot F_{ig} \cdot B'_o + A'_o \cdot F_{ig} \cdot B'_i + A'_o \cdot G_{sv} \cdot R_{eg}}$ |

| Closed-loop audio-susceptibility transfer function, $G_{vv}(s)$ |
|---|
| $$\frac{\hat{v}_o}{\hat{v}_g} = \frac{C'_o + A'_o \cdot F_{vg} - A'_i \cdot F_{ig} \cdot C'_o + A'_o \cdot F_{ig} \cdot C'_i}{B'_o - \frac{A'_o \cdot F_{io}}{Z_{load}} - \frac{A'_i \cdot F_{ig}}{Z_{load}} - A'_i \cdot B'_o \cdot F_{ig} + A'_o \cdot B'_i \cdot F_{ig} + A'_o \cdot G_{sv} \cdot R_{eg} + \frac{1}{Z_{load}}}$$ |
| Unterminated closed-loop back-current transfer function, $G_{iio}(s)$ |
| $$G_{iio}(s) = \frac{\hat{\imath}_g}{\hat{\imath}_o} = \frac{B'_i + A'_i \cdot B'_o \cdot F_{io} - A'_o \cdot B'_i \cdot F_{io} + A'_i \cdot G_{sv} \cdot R_{eg}}{B'_o - A'_i \cdot B'_o \cdot F_{ig} + A'_o \cdot B'_i \cdot F_{ig} + A'_o \cdot G_{sv} \cdot R_{eg}}$$ |

The open-loop transfer functions can be deduced from closed-loop transfer functions by setting to zero the term $R_{eg}(s)$. An original contribution of this paper is the deduction of the converter transfer functions and impedances that are shown in Table V.

In addition, the methodology proposed in this paper can be easily scalable to other types of input filters such as CLC as shown in Fig.10. In this new scenario it is necessary to replace the term $C'_i(s)$ by $C''_i(s)$ in the converter transfer functions that were obtained in Table V. The characteristic coefficient $C''_i(s)$ is given by (22). Therefore, adding new types of filters may not lead to having to carry out the whole process of the proposed methodology again.

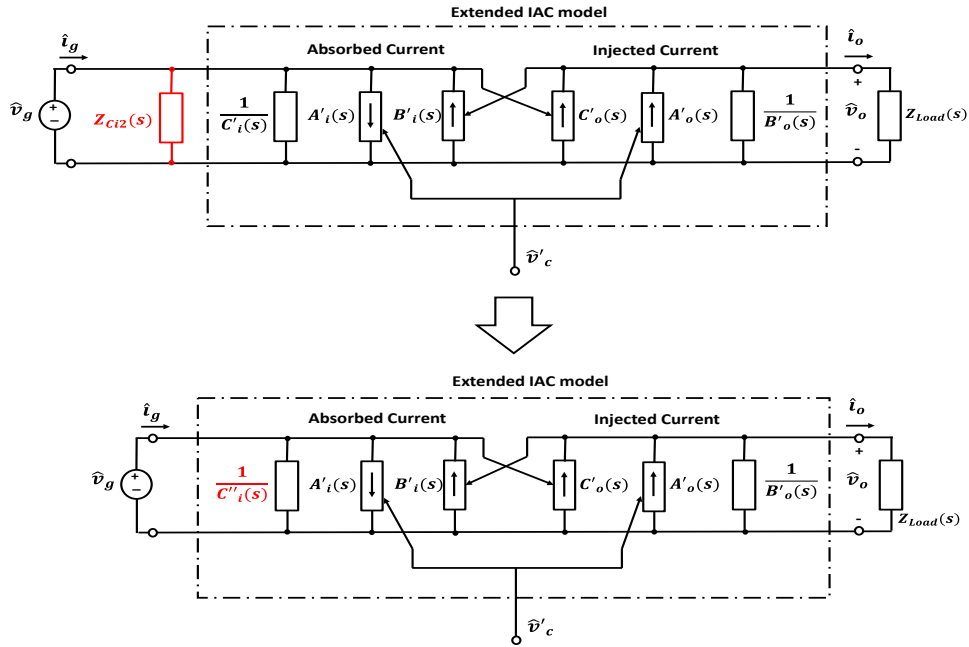

Fig. 10 EIAC method applied to other input filter configuration

$$C''_i(s) = \frac{1}{Z_{Ci2}(s) \parallel \frac{1}{C'_i(s)}} \qquad (22)$$

As a summary, Fig. 11 shows a diagram that indicates how to use the transfer functions for the different control and power structures.

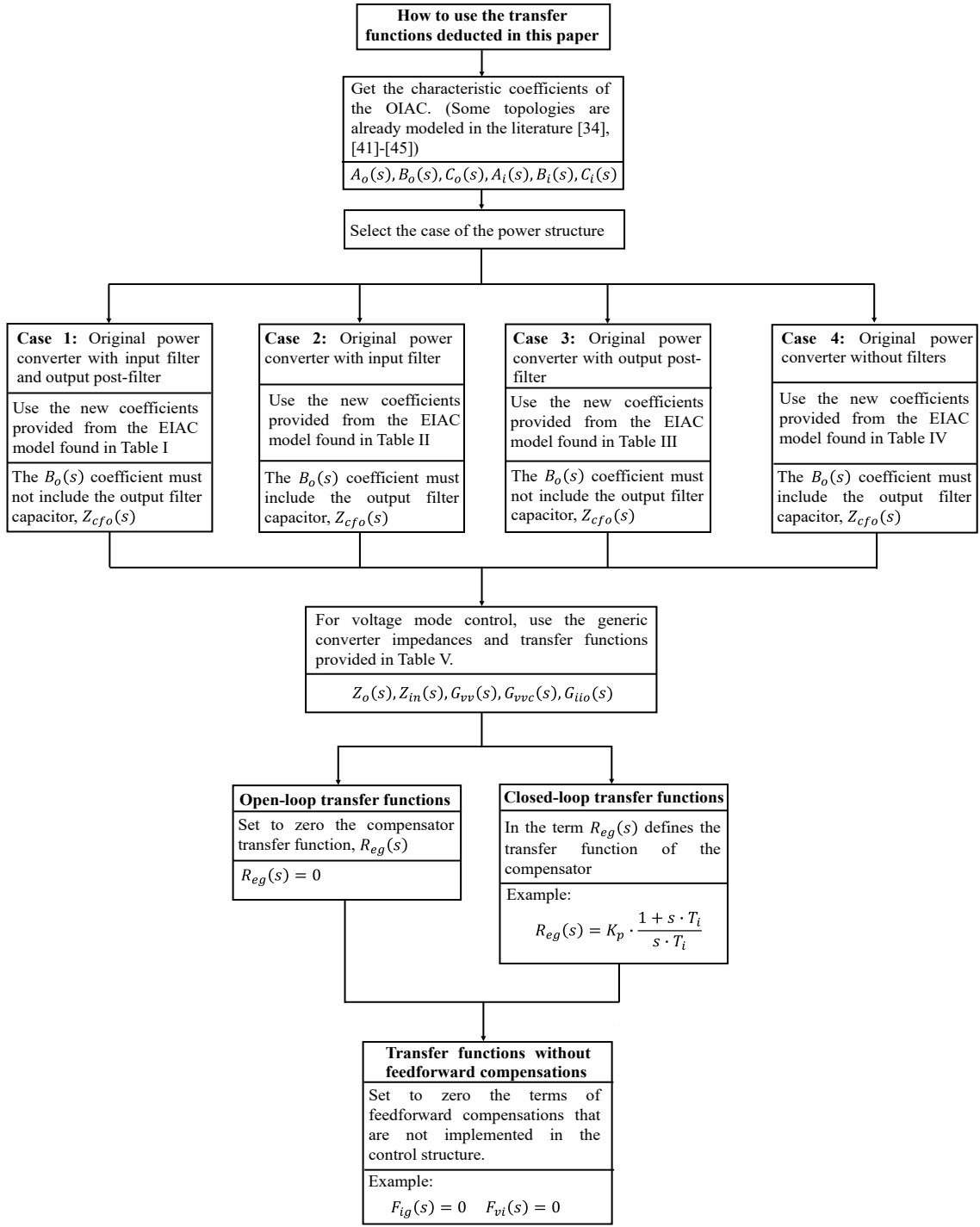

Fig. 11 Diagram that indicates how to use the transfer functions

Finally, as an example, to obtain the equivalent transfer function of the open-loop output impedance for a power converter with input filter, output post-filter and feedforward compensation of the output current, the characteristic coefficients that are in Table I should be used. The terms $F_{ii}$ and $F_{vi}$ must be set to zero, since there are no implementations of the feedforward compensations of the input port. Then the transfer function of the closed-loop output impedance is chosen from Table V and the terms $F_{ig}$ and $F_{vg}$ must also be set to zero. Finally, in order to obtain the open-loop transfer function, the compensator transfer function, $R_{eg}(s)$, must be set to zero.

In conclusion, having a set of generic transfer functions allows to quickly design the control loop, since these transfer functions can be programmed by using different types of mathematical programs such as Mathcad or MATLAB, etc.

## III. VALIDATION OF THE EXTENSION OF THE IAC METHOD

The validation of the extension of the IAC method has been carried out experimentally and simulated. For the experimental validation, two prototypes of Phase-Shifted Full-Bridge converter were built. The first prototype was designed to withstand a maximum power of 20-kW and the second prototype to withstand a maximum power of 200-W.

### A. Experimental measurements on the 20-kW prototype

This prototype has been used to validate the transfer functions of power structure 2. The values of the components of the schematic shown in Fig. 12 are listed in Table VI. This converter is controlled by Zybo board [54] based on 7Z010 Xilinx Zynq device [55] as shown in Fig. 13. The Full-Bridge is composed of two SiC power modules (CAS300M17BM2) from CREE company [56]. The schematic of the analog sensing chain used in the closed-loop transfer functions is shown in Fig. 14. The frequency response is obtaining by using the Venable frequency response analyzer model 3235 [57].

TABLE VI

20-kW PROTOTYPE COMPONENT VALUES

| $L_i = 30\ mH$ | $C_{if\_1} = 220\ \mu F$ | $C_{if\_1} = 220\ \mu F$ | $R_{i\_1} = 82\ k\Omega$ |
| --- | --- | --- | --- |
| $R_{i\_1} = 82\ k\Omega$ | $R_{i\_3} = 24\ k\Omega$ | $R_d = 2.2\ \Omega$ | $C_d = 330\ nF$ |
| $L_{lk} = 5\ \mu H$ | $n = 0.471$ | $L = 130\ \mu H$ | $C_{fo\_1} = 6800\ \mu H$ |
| $C_{fo\_2} = 6800\ \mu H$ | $C_{fo\_3} = 6800\ \mu H$ | $F_{sw} = 10\ kHz$ | $R_{load} = 1.76\ \Omega$ |

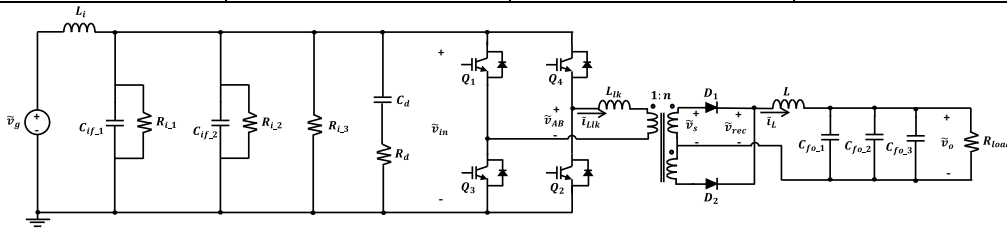

Fig. 12 Schematic of 20-kW prototype

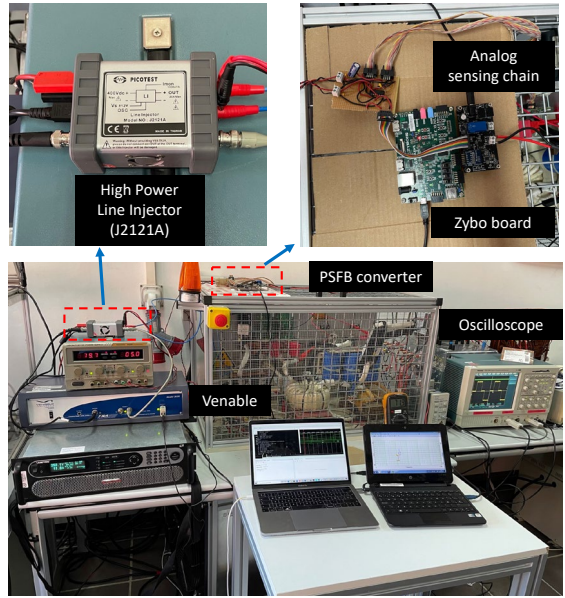

Fig. 13 20-kW Phase-Shifted Full-Bridge prototype

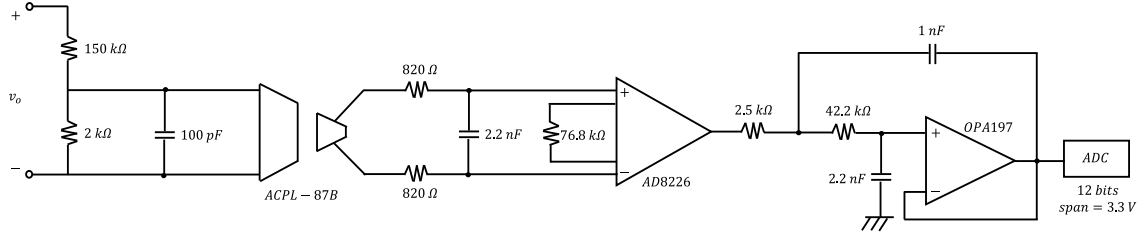

Fig. 14 Schematic of the analog sensing chain

The first transfer function to be validated is the control to output voltage, $G_{vvc}(s)$. The input voltage, $v_g$, is equal to 250V and the output voltage, $v_o$, is equal to 72V. The main waveforms of this operating point are shown in Fig. 15. Although the converter supports a power of 20-kW due to the load limitation, the maximum power obtained in the experimental measurement is approximately 3-kW.

None of the FF compensations have been implemented in the control structure of this prototype. So, the terms $F_{vi}(s)$, $F_{ii}(s)$, $F_{vg}(s)$, $F_{ig}(s)$ and $F_{io}(s)$ are set to 0.

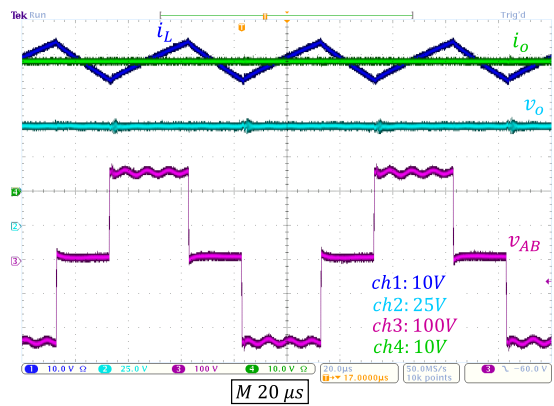

Fig. 15 Main waveforms of the 20-kW prototype

The measurement block diagram to obtain the frequency response of the control to output voltage transfer function is shown in Fig. 16.

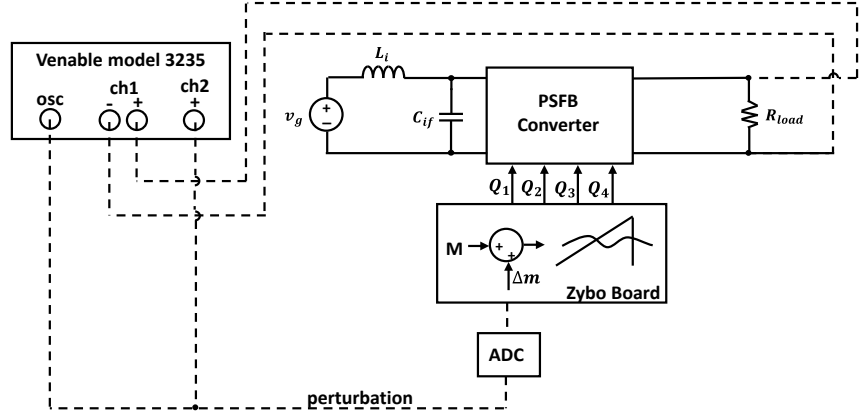

Fig. 16 Measurement block diagram of the frequency response of the control to output voltage

The oscillator signal, $osc$, from the Venable is sampled through an ADC with a sample rate equal to $2 \cdot F_{sw}$. This sampling frequency has been selected because the ripple of the secondary waveforms is twice the switching frequency, as can be seen in Fig. 15. The time diagram of the digital control implemented in this prototype is shown in Fig. 17.

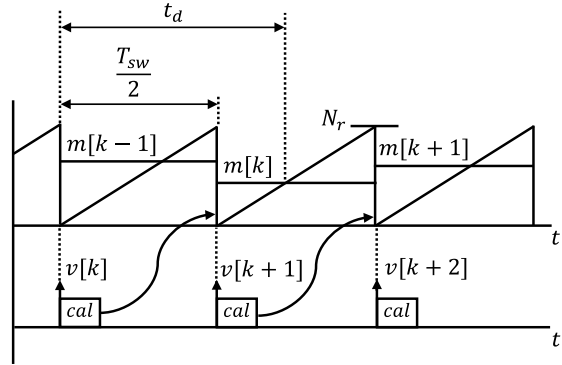

Fig. 17 Timing diagram of the digital control

The modulator transfer function is defined by (23).

$$G_m(s) = \frac{1}{N_r} \cdot e^{-s \cdot t_d} \quad (23)$$

Where $N_r$ is the amplitude of the sawtooth carrier, $m$ represents the modulating signal and $t_d$ represents the transport delay [58].

$$t_d = \frac{T_{sw}}{2} + D \cdot \frac{T_{sw}}{2} \quad (24)$$

The characteristic coefficients of the OIAC method that are considered for the validation are obtained from the small-signal model of the Phase-Shifted Full-Bridge converter presented in [45].

The frequency response of the control to output voltage transfer function when the converter has no input filter is shown in Fig. 18(a). It can be seen from Fig. 18(a) that the experimental measurement matches with the theoretical model. The phase drop that occurs around 1 kHz is due to the effect of the transport delay. On the other hand, Fig. 18(b) shows the frequency response of the control to output voltage transfer function when the converter has an input filter. From Fig. 18(b), it be seen how the input filter modifies the magnitude and phase in the vicinity of the input filter resonance frequency ($F_r = 40\ Hz$). The resonant frequency of the input filter is given by (25). The phase in Fig. 18(b) is also affected by the transport delay.

$$F_r = \frac{1}{2 \cdot \pi \cdot \sqrt{L_i \cdot C_{if}}} \tag{25}$$

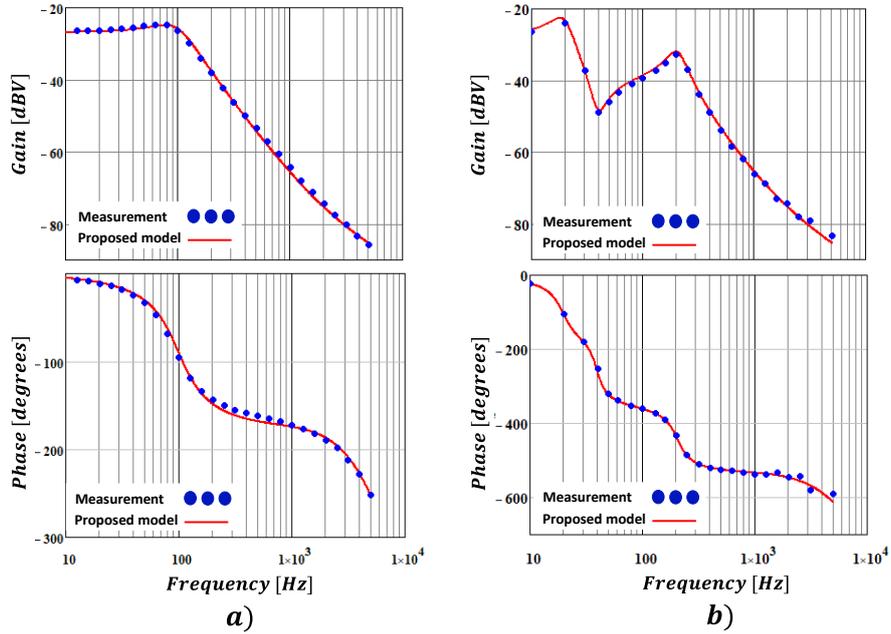

Fig. 18 Comparison between experimental frequency response and theoretical frequency response of control to output voltage transfer function ($v_g = 250V;\ v_o = 72V;\ D = 0.6$): a) without input filter b) with input filter

For the closed-loop transfer functions, the compensator coefficients have been tuned using the SmartCtrl software [59]. Fig. 19(a) and Fig. 19(b) show the characteristics of the compensator designed for the case when the converter has no input filter and when the converter has an input filter, respectively. The open-loop transfer function, $T(s)$, is defined by (26).

$$T(s) = G_{vvc}(s) \cdot G_{adc}(s) \cdot G_{sv}(s) \cdot R_{eg}(s) \tag{26}$$

The term $G_{sv}(s)$ represents the transfer function of the sensing stage shown in Fig. 14 and the term $G_{ADC}(s)$ represents the analog-digital-converter transfer function. From Fig. 19(b) it can be seen that the bandwidth of the compensator is limited by the resonance frequency of the input filter, for this reason it has been necessary to reduce the cutoff frequency, because if a cutoff frequency of 131 Hz is selected, the phase would be below -180º and the loop would become unstable. This analysis demonstrates the need to have transfer functions that take into account the input filter.

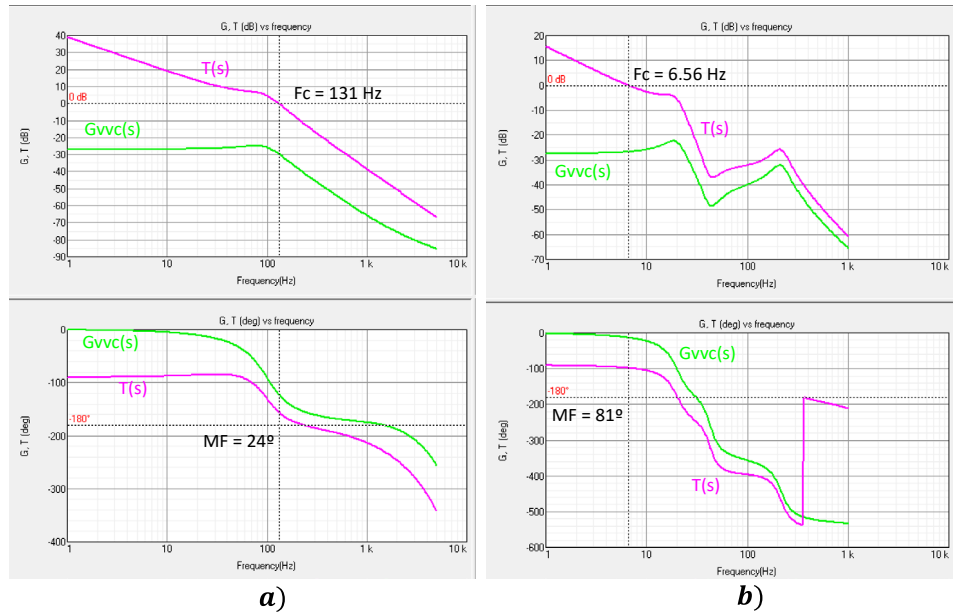

Fig. 19 Compensator design: a) converter without input filter (Kp = 1 and Ti = 0.0022) b) converter with input filter (Kp = 0.073 and Ti = 0.0022)

The second transfer function to be validated is the closed-loop audio-susceptibility, $G_{vv}(s)$. The measurement block diagram to obtain the frequency response of the closed-loop audio-susceptibility transfer function is shown in Fig. 20. The perturbation in the input voltage, $v_g$, is injected using High Power Line Injector J2121A from Picotest company [60] as shown in Fig. 13 and Fig. 20.

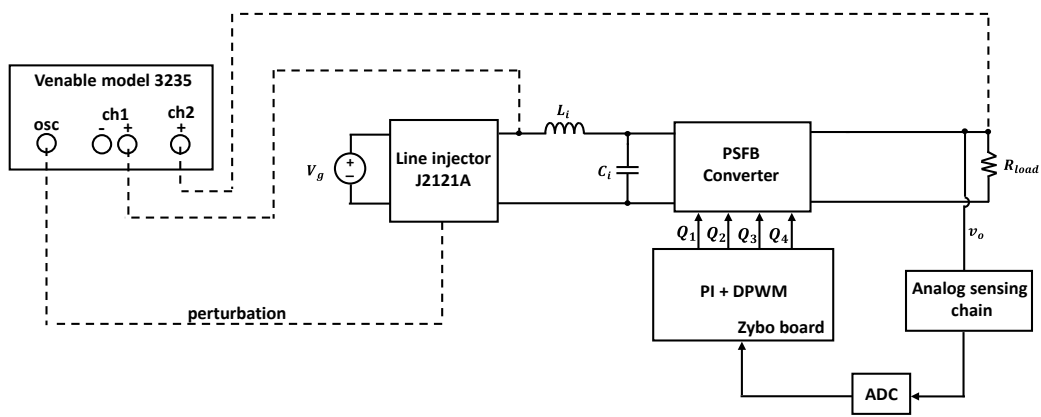

Fig. 20 Measurement block diagram of the frequency response of the closed-loop audio-susceptibility

The frequency response of the closed-loop audio-susceptibility transfer function when the converter has no input filter is shown in Fig. 21(a). On the other hand, Fig. 21(b) shows the frequency response of the closed-loop audio-susceptibility transfer function when the converter has an input filter. Fig. 21(b) shows how both the input filter and the transport delay cause an additional phase drop. The experimental measurement matches with the theoretical model.

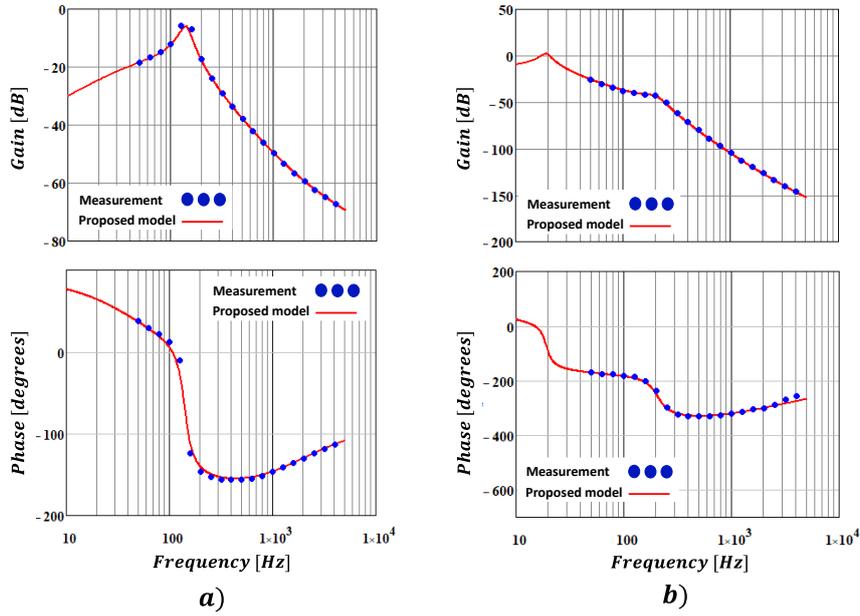

Fig. 21 Comparison between experimental frequency response and theoretical frequency response of closed-loop audio-susceptibility transfer function: a) without input filter (Kp = 1 and Ti = 0.0022) b) with input filter (Kp = 0.073 m and Ti = 0.0022)

Finally, the frequency response of the closed-loop input impedance, $Z_{in}(s)$, of the converter is obtained. The measurement block diagram is shown in Fig. 22. Again, the input voltage, $v_g$, is perturbed with the line injector J2121A. The measuring points are input current, $i_g$, and input voltage, $v_g$. The input capacitors have been considered for this measurement.

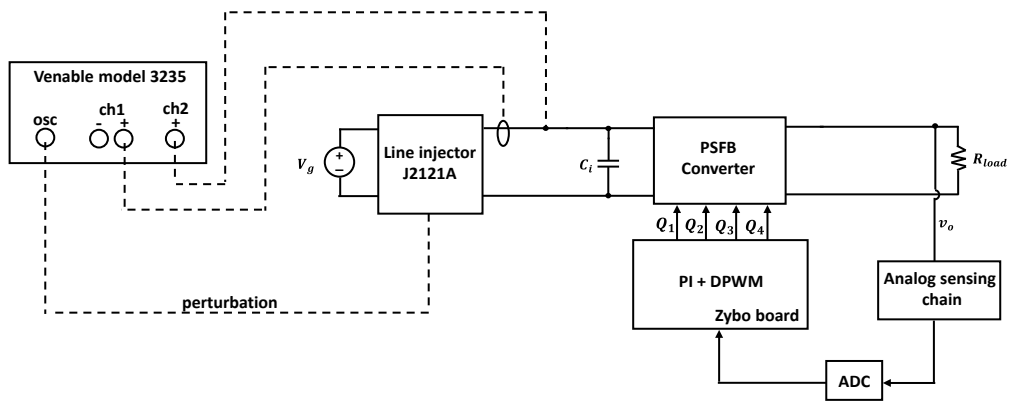

Fig. 22 Measurement block diagram of the frequency response of the closed-loop input impedance

The frequency response of the closed-loop input impedance transfer function is shown in Fig. 23. It can be seen from Fig. 23 that the experimental measurement matches with the theoretical model. At low frequency, the constant power load effect of the converter can be observed.

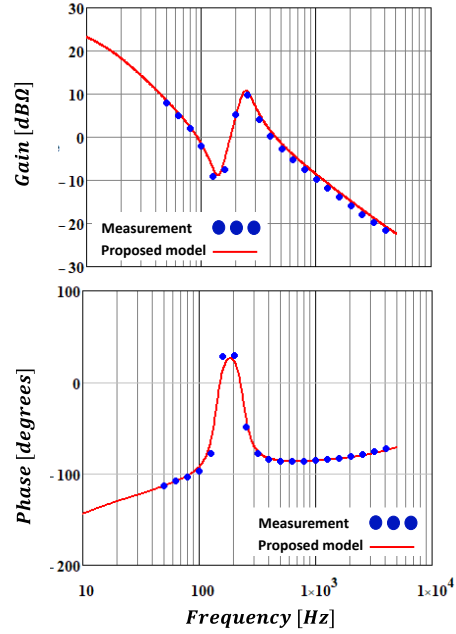

Fig. 23 Comparison between experimental frequency response and theoretical frequency response of closed-loop input impedance transfer function without input filter (Kp = 1 and Ti = 0.0022)

B. Experimental measurements on the 200-W prototype

This prototype has been used to validate the transfer functions of power structure 1. The schematic of the second prototype is shown in Fig. 24. The values of the components are listed in Table VI. This converter is also controlled by the Zybo board as shown in Fig. 25. The main waveforms of the operating point are shown in Fig. 26.

TABLE VII

200 W PROTOTYPE COMPONENT VALUES

| $L_i = 38\ mH$ | $C_{if} = 100\ \mu F$ | $L_{lk} = 5\ \mu H$ | $n = 0.5$ |
|---|---|---|---|
| $L = 36\ \mu H$ | $C_{fo} = 47\ \mu F$ | $L_p = 10\ \mu H$ | $C_p = 22\ \mu F$ |
| $F_{sw} = 100\ kHz$ | $R_{load} = 2.2\ \Omega$ | $V_o = 20\ V$ | $V_g = 100\ V$ |

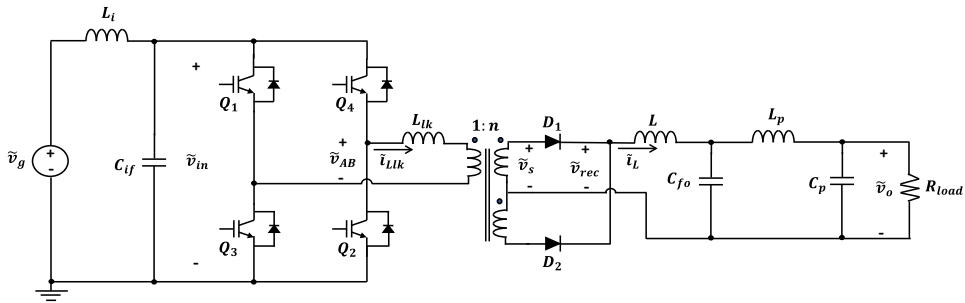

Fig. 24 Schematic of 200-W PSFB converter

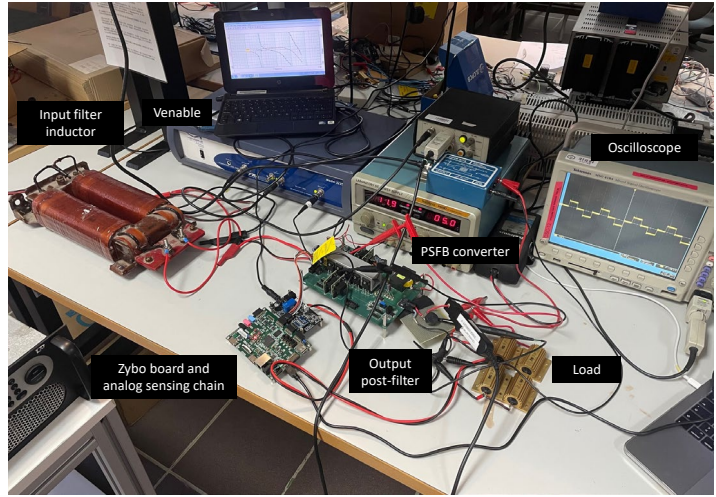

Fig. 25 200-W Phase-Shifted Full-Bridge prototype

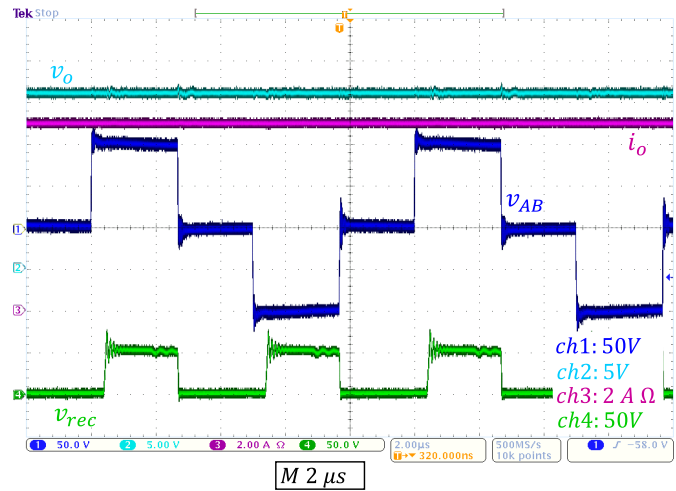

Fig. 26 Main waveforms of the 200-W prototype

The first transfer function to be validated is the control to output voltage, $G_{vvc}(s)$. The Measurement block diagram of this transfer function is the same as shown in Fig.16. Therefore, the transfer function of the modulator is defined by (23). The comparison between the experimental frequency response and the theoretical frequency response is shown in Fig. 27(a). In this case, none Feedforward compensations have been implemented. So, the $F_{vi}(s)$, $F_{ii}(s)$, $F_{vg}(s)$, $F_{ig}(s)$ and $F_{io}(s)$ are set to 0. The resonant frequency of the input filter is approximately 81 Hz. For this reason, the first phase drop can be observed in the vicinity of the resonant frequency. A second phase drop occurs around 10 kHz due to the output post-filter and transport delay. In Fig. 27(b) the frequency response has been obtained by simulation using the "Ac sweep" function of the PSIM software [61]. In this case the feedforward compensation of the output current $(F_{io}(s) = 10)$ has been taken in account. The other terms of the FF compensations are set to 0. In this transfer function the $F_{io}(s)$ term has been adjusted in order to show the effect it has on the transfer function.

Fig. 27(b) shows how the FF compensation can modify the magnitude and phase. The experimental and simulated frequency responses match with the theoretical responses.

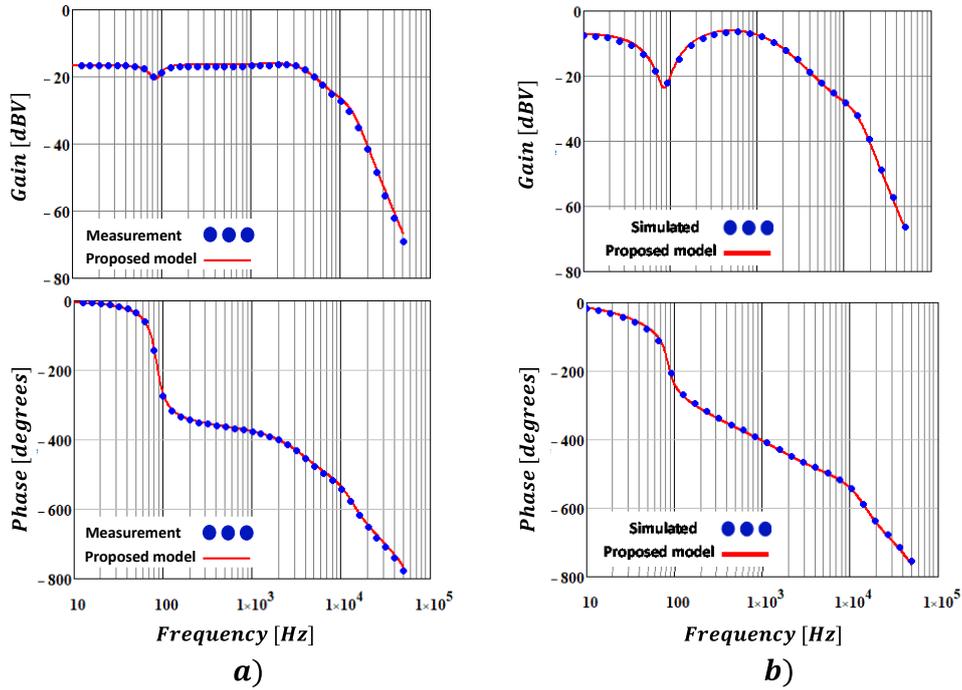

Fig. 27 Frequency response comparison of the control to output voltage transfer function: a) without implementing FF compensations b) considering the FF compensation of the output current $(F_{io}(s) = 10)$

The next transfer function to be validated is the closed-loop input impedance. The Measurement block diagram of this transfer function is the same as shown in Fig. 22. The input filter inductor has been removed for this validation. This measurement considers the converter input capacitor, $C_{if}$. The sensing stage for the closed-loop transfer functions is the same circuit that was used in the 20-kW prototype, so the circuit shown in Fig.14 is valid for the 200-W prototype. The comparison between the experimental frequency response and the theoretical frequency response is shown in Fig. 28(a). In this case, none Feedforward compensations have been implemented. So, the $F_{vi}(s)$, $F_{ii}(s)$, $F_{vg}(s)$, $F_{ig}(s)$ and $F_{io}(s)$ are set to 0. It can be seen from Fig. 28(a) that at low frequency the constant power load behavior as the phase tends to start from -180º. In Fig. 28(b) the frequency response has been obtained by simulation using the "Ac sweep" function of the PSIM software. In this case the feedforward compensation of the input voltage $(F_{vg}(s) = 10)$ has been taken in account. The other FF compensations are set to 0. In this transfer function the $F_{vg}(s)$ term has been adjusted in order to show the effect it has on the transfer function. Fig. 28(b) shows how the FF compensation can modify the magnitude and phase. The experimental and simulated frequency responses match with the theoretical responses.

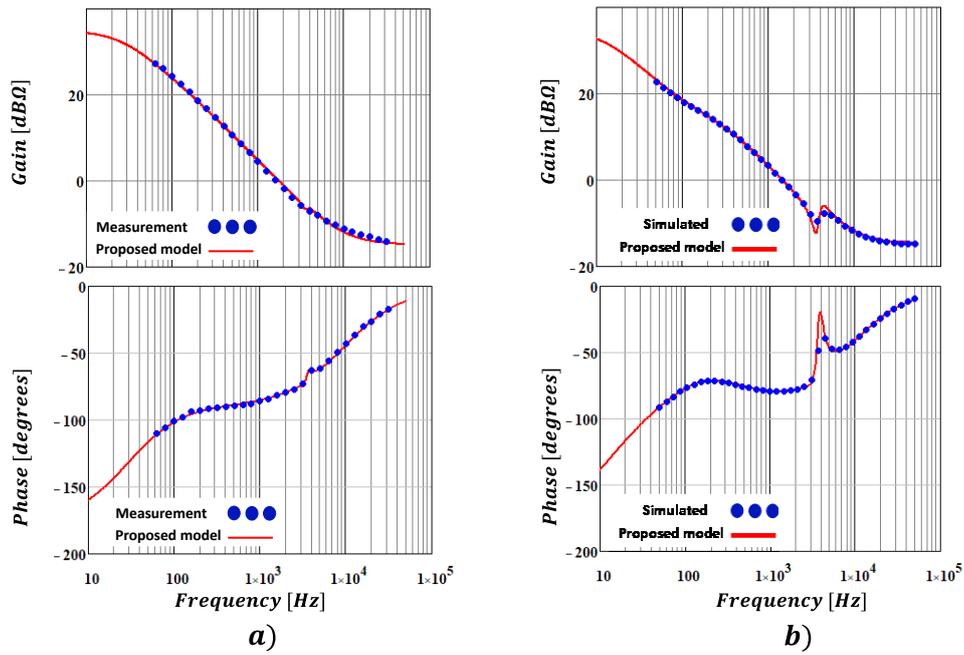

Fig. 28 Closed-loop input impedance comparison: a) without implementing FF compensations (Kp = 0.5 and Ti = 0.001) b) considering the FF compensation of the input voltage $\left(F_{vg}(s) = 10\right)$

The comparison between the experimental frequency response and the theoretical frequency response of the closed-loop audio-susceptibility is shown in Fig. 29. In this case, none Feedforward compensations have been implemented. So, the terms $F_{vi}(s)$, $F_{ii}(s)$, $F_{vg}(s)$, $F_{ig}(s)$ and $F_{io}(s)$ are set to 0. The Measurement block diagram of this transfer function is the same as shown in Fig. 20. The experimental and simulated frequency response match with the theoretical responses.

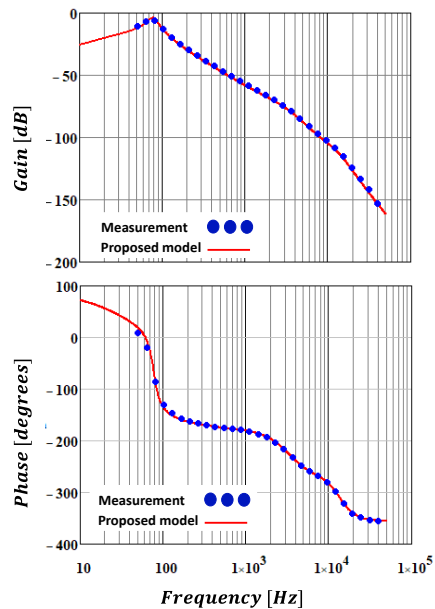

Fig. 29 Comparison between experimental frequency response and theoretical frequency response of closed-loop audio-susceptibility transfer function (Kp = 0.05 and Ti = 0.001)

Finally, the closed-loop output impedance and back-current transfer function are validated by simulation, as shown in Fig. 30(a) and Fig. 30(b), respectively. Both transfer functions match with the proposed model. Fig. 30(a) and Fig. 30(b) show how the input filter abruptly affects the magnitude and phase in the vicinity of the resonant frequency ($F_r = 81\ Hz$). In the case that a cascaded converter is connected then the resonance peak of the input filter is very important to consider in the output impedance since in that scenario the analysis of the minor loop gain, $TMLG$, must be used in order to ensure the stability of the whole system. On the other hand, the output post-filter produces a second phase drop around 10kHz.

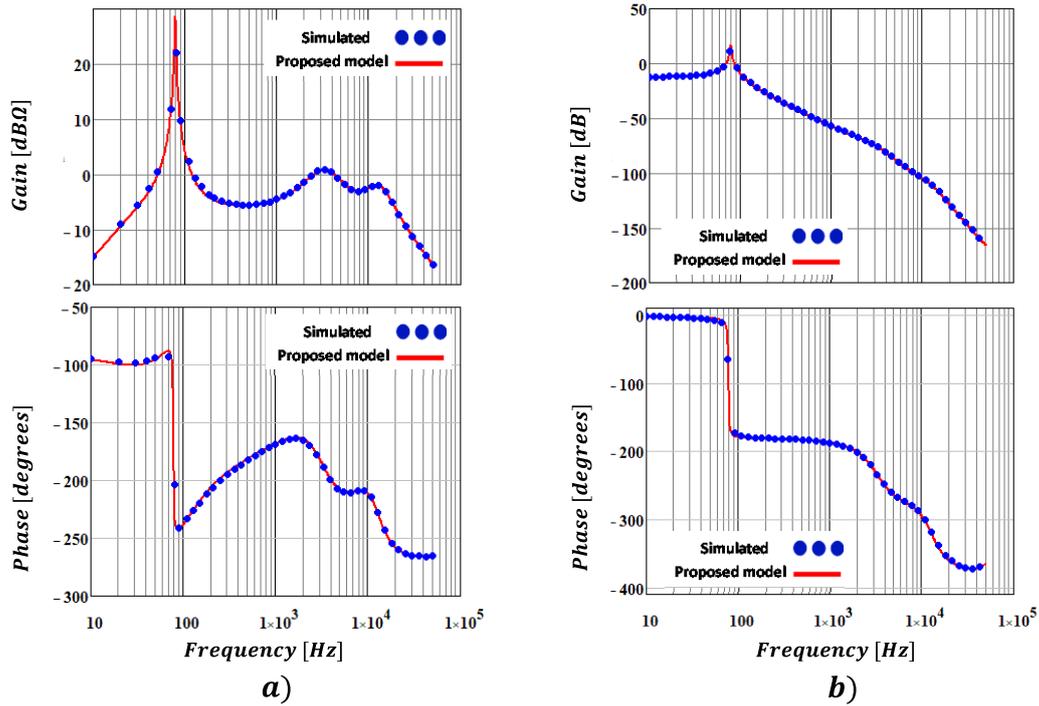

Fig. 30 Frequency response comparison: a) closed-loop output impedance without implementing FF compensations (Kp = 0.05 and Ti = 0.001) b) closed-loop back-current transfer function without implementing FF compensations (Kp = 0.05 and Ti = 0.001)

It has been demonstrated that the proposed methodology allows predicting the frequency response of different power and control structures. The EIAC model can be very useful for the analysis of the stability of auxiliary systems in railway applications.

IV. CONCLUSIONS

In this paper a methodology has been proposed that allows extending the injected-absorbed-current method to cases where a converter may have an input filter or an output post-filter or both filters at the same time. In addition, the resulting model allows to take into account five different feedforward compensations. The main advantage of the EIAC method is that it allows reusing the characteristic coefficients of the DC-DC converter derived from the IAC method. The transfer functions that have been obtained in previous works from the IAC method can be reusable in the case of the EIAC model thanks to the similarity that exist between both equivalent circuit.

It has been shown that the EIAC method can consider different types of loads such as resistive load, constant current load and constant power load, CPL. The CPL are very common to be found in the converters of the auxiliary systems of railway applications. Therefore, it has been demonstrated that in railway applications the need to determine transfer functions such as input and output impedance since the minor loop gain, TMLG, is defined by the impedance ratio. The TMLG is responsible for system stability when 2 converters are connected in cascade or in the case of a converter with an input filter.

On the other hand, the EIAC model have been validated by simulation and experimentally. For the experimental validation has been built two prototypes. The 20-kW prototype was used to validate the transfer functions of power structure 2 (converter with input filter) and the 200-W prototype was used to validate the transfer functions of power structure 1 (converter with input filter and output post-filter). It has been demonstrated that the input filter and the output post-filter affect the magnitude and phase of all the transfer functions obtained in this paper.

Finally, from the experimental validations it could be observed that the cut-off frequency of the compensator is limited by the resonance frequency of the input filter since in railway applications the resonance frequency of this filter is generally very low.

# APPENDIX A

The new characteristic coefficients are functions of the complex variable "s". In this analysis the dependence on the Laplace variable has been omitted for simplicity.

**Deduction of the control to output voltage transfer function**

If (19) is substituted into (5) then the input current, $\hat{\imath}_g$, is given by (a.1).

$$\hat{\imath}_g = \frac{A'_i}{1 - A'_i \cdot F_{ig}} \cdot \hat{v}_c + \frac{A'_i \cdot F_{io}}{1 - A'_i \cdot F_{ig}} \cdot \hat{\imath}_o + \frac{A'_i \cdot F_{vg} + C'_i}{1 - A'_i \cdot F_{ig}} \cdot \hat{v}_g - \frac{B'_i}{1 - A'_i \cdot F_{ig}} \cdot \hat{v}_o \tag{a.1}$$

If $v_g$ is a constant voltage, then $\hat{v}_g = 0$.

$$\hat{\imath}_g = \frac{A'_i}{1 - A'_i \cdot F_{ig}} \cdot \hat{v}_c + \frac{A'_i \cdot F_{io}}{1 - A'_i \cdot F_{ig}} \cdot \hat{\imath}_o - \frac{B'_i}{1 - A'_i \cdot F_{ig}} \cdot \hat{v}_o \tag{a.2}$$

Substituting (20) in (a.2)

$$\hat{\imath}_g = \frac{A'_i}{1 - A'_i \cdot F_{ig}} \cdot \hat{v}_c + \left(\frac{A'_i \cdot F_{io}}{Z_{load} \cdot (1 - A'_i \cdot F_{ig})} - \frac{B'_i}{1 - A'_i \cdot F_{ig}}\right) \cdot \hat{v}_o \tag{a.3}$$

On the other hand, substituting (19) and (20) in (6).

$$\frac{\hat{v}_o}{Z_{load}} = A'_o \cdot \hat{v}_c + \left(\frac{A'_o \cdot F_{io}}{Z_{load}} - B'_o\right) \cdot \hat{v}_o + A'_o \cdot F_{ig} \cdot \hat{\imath}_g \tag{a.4}$$

The control to output voltage transfer function is obtained by substituting (a.3) in (a.4).

$$\hat{v}_o = \frac{A'_o \cdot Z_{load}}{F_{ig} \cdot Z_{load} \cdot (A'_o \cdot B'_i - A'_i \cdot B'_o) + 1 + B'_o \cdot Z_{load} - A'_o \cdot F_{io} - A'_i \cdot F_{ig}} \cdot \hat{v}_c \tag{a.5}$$

**Deduction of the closed-loop input impedance**

If (20) and (21) are substituted into (6) then the output voltage, $\hat{v}_o$, is given by (a.6).

$$\hat{v}_o = \frac{A'_o \cdot F_{vg} + C'_o}{\frac{1}{Z_{load}} + A'_o \cdot G_{sv} \cdot R_{eg} - \frac{A'_o \cdot F_{io}}{Z_{load}} + B'_o} \cdot \hat{v}_g + \frac{A'_o \cdot F_{ig}}{\frac{1}{Z_{load}} + A'_o \cdot G_{sv} \cdot R_{eg} - \frac{A'_o \cdot F_{io}}{Z_{load}} + B'_o} \cdot \hat{\imath}_g \tag{a.6}$$

Substituting (20) and (21) in (5) gives the input current, $\hat{\imath}_g$, as a function of the output voltage, $\hat{v}_o$, and the input voltage, $\hat{v}_g$.

$$\hat{\imath}_g = \frac{\frac{A'_i \cdot F_{io}}{Z_{load}} - A'_i \cdot G_{sv} \cdot R_{eg} - B'_i}{1 - A'_i \cdot F_{ig}} \cdot \hat{v}_o + \frac{A'_i \cdot F_{vg} + C'_i}{1 - A'_i \cdot F_{ig}} \cdot \hat{v}_g \tag{a.7}$$

The closed-loop input impedance is obtained by substituting (a.6) in (a.7).

$$\frac{\hat{v}_g}{\hat{\imath}_g} = \frac{B'_o - \frac{A'_o \cdot F_{io}}{Z_{load}} - \frac{A'_i \cdot F_{ig}}{Z_{load}} - A'_i \cdot B'_o \cdot F_{ig} + A'_o \cdot B'_i \cdot F_{ig} + A'_o \cdot G_{sv} \cdot R_{eg} + \frac{1}{Z_{load}}}{\frac{C'_i}{Z_{load}} + \frac{A'_i \cdot F_{vg}}{Z_{load}} - C'_o \cdot B'_i + C'_i \cdot B'_o + \frac{F_{io}}{Z_{load}} \cdot (A'_i \cdot C'_o - A'_o \cdot C'_i) + F_{vg} \cdot (A'_i \cdot B'_o - A'_o \cdot B'_i) - G_{sv} \cdot R_{eg} \cdot (A'_i \cdot C'_o - A'_o \cdot C'_i)} \tag{a.8}$$

**Deduction of unterminated closed-loop output impedance**

If (21) is substituted into (5) then the input current, $\hat{\imath}_g$, is given by (a.9).

$$\hat{\imath}_g = \frac{A'_i \cdot F_{io}}{1 - A'_i \cdot F_{ig}} \cdot \hat{\imath}_o - \frac{A'_i \cdot G_{sv} \cdot R_{eg} + B'_i}{1 - A'_i \cdot F_{ig}} \cdot \hat{v}_o + \frac{A'_i \cdot F_{vg} + C'_i}{1 - A'_i \cdot F_{ig}} \cdot \hat{v}_g \qquad (a.9)$$

If $v_g$ is a constant voltage, then $\hat{v}_g = 0$.

$$\hat{\imath}_g = \frac{A'_i \cdot F_{io}}{1 - A'_i \cdot F_{ig}} \cdot \hat{\imath}_o - \frac{A'_i \cdot G_{sv} \cdot R_{eg} + B'_i}{1 - A'_i \cdot F_{ig}} \cdot \hat{v}_o \qquad (a.10)$$

Substituting (21) in (6) gives the output current, $\hat{\imath}_o$, as a function of the output voltage, $\hat{v}_o$, and the input current, $\hat{\imath}_g$.

$$\hat{\imath}_o = \frac{F_{ig} \cdot A'_o}{1 - A'_o \cdot F_{io}} \cdot \hat{\imath}_g - \frac{B'_o + A'_o \cdot G_{sv} \cdot R_{eg}}{1 - A'_o \cdot F_{io}} \cdot \hat{v}_o \qquad (a.11)$$

The unterminated closed-loop output impedance is obtained by substituting (a.10) in (a.11).

$$\frac{\hat{v}_o}{\hat{\imath}_o} = \frac{A'_i \cdot F_{ig} + A'_o \cdot F_{io} - 1}{B'_o - A'_i \cdot F_{ig} \cdot B'_o + A'_o \cdot F_{ig} \cdot B'_i + A'_o \cdot G_{sv} \cdot R_{eg}} \qquad (a.12)$$

**Deduction of closed-loop audio-susceptibility transfer function**

If (20) and (21) are substituted into (6) then the output voltage, $\hat{v}_o$, is given by (a.13).

$$\hat{v}_o = \frac{A_o \cdot F_{vg} + C_o}{\frac{1}{Z_{load}} + A_o \cdot G_{sv} \cdot R_{eg} - \frac{A_o \cdot F_{io}}{Z_{load}} + B_o} \cdot \hat{v}_g + \frac{A_o \cdot F_{ig}}{\frac{1}{Z_{load}} + A_o \cdot G_{sv} \cdot R_{eg} - \frac{A_o \cdot F_{io}}{Z_{load}} + B_o} \cdot \hat{\imath}_g \qquad (a.13)$$

The closed-loop audio-susceptibility is obtained by substituting (a.7) in (a.13).

$$\frac{\hat{v}_o}{\hat{v}_g} = \frac{C'_o + A'_o \cdot F_{vg} - A'_i \cdot F_{ig} \cdot C'_o + A'_o \cdot F_{ig} \cdot C'_i}{B'_o - \frac{A'_o \cdot F_{io}}{Z_{load}} - \frac{A'_i \cdot F_{ig}}{Z_{load}} - A'_i \cdot B'_o \cdot F_{ig} + A'_o \cdot B'_i \cdot F_{ig} + A'_o \cdot G_{sv} \cdot R_{eg} + \frac{1}{Z_{load}}} \qquad (a.14)$$

**Deduction of unterminated closed-loop back current transfer function**

If it is considered that the voltage $v_g$ is kept constant, then the expression (a.10) is valid for this transfer function. From (a.10) it can be inferred that the output voltage, $\hat{v}_o$, is given by (a.15).

$$\hat{v}_o = \frac{A'_i \cdot F_{io}}{A'_i \cdot G_{sv} \cdot R_{eg} + B'_i} \cdot \hat{\imath}_o - \frac{1 - A'_i \cdot F_{ig}}{A'_i \cdot G_{sv} \cdot R_{eg} + B'_i} \cdot \hat{\imath}_g \qquad (a.15)$$

Since the expression (a.11) is also valid for this transfer function, then it can be inferred that the input current, $\hat{\imath}_g$, is given by (a.16).

$$\hat{\imath}_g = \frac{1 - A'_o \cdot F_{io}}{F_{ig} \cdot A'_o} \cdot \hat{\imath}_o + \frac{B'_o + A'_o \cdot G_{sv} \cdot R_{eg}}{F_{ig} \cdot A'_o} \cdot \hat{v}_o \qquad (a.16)$$

The unterminated closed-loop back-current transfer function is obtained by substituting (a.15) in (a.16).

$$\frac{\hat{\imath}_g}{\hat{\imath}_o} = \frac{B'_i + A'_i \cdot B'_o \cdot F_{io} - A'_o \cdot B'_i \cdot F_{io} + A'_i \cdot G_{sv} \cdot R_{eg}}{B'_o - A'_i \cdot B'_o \cdot F_{ig} + A'_o \cdot B'_i \cdot F_{ig} + A'_o \cdot G_{sv} \cdot R_{eg}} \qquad (a.17)$$